\documentclass[a4paper,12pt]{article}
\usepackage[dvips]{graphicx}
\begin{document}
\hyphenation{extra-po-lation Re-la-ti-vi-stic re-la-ti-vi-stic}
\begin{center}
\Large\bf
From RHIC to LHC: A relativistic diffusion approach\\[2.1cm]
\large\rm
Rolf Kuiper and Georg Wolschin
\footnote{Corresponding author. Fax:+49-6221-549333.
Email: wolschin@uni-hd.de}\\[.8cm]
\normalsize\sc\rm
Institut f\"ur Theoretische Physik der Universit\"at Heidelberg\\
Philosophenweg 16,
D-69120 Heidelberg, Germany\\[2.6cm]
%(Submitted July 28, 2006)\\[2.0cm]
\end{center}
\rm
We investigate the energy dependence of stopping and
hadron production in high-energy heavy-ion collisions
based on a three-sources Relativistic Diffusion Model. The
transport coefficients are extrapolated from Au + Au
and Cu + Cu at RHIC energies
($\sqrt{s_{NN}}$=19.6 - 200 GeV) to Pb + Pb at LHC energies
$\sqrt{s_{NN}}$= 5.52 TeV. Rapidity distributions for net protons,
and pseudorapidity spectra for produced charged particles in central
collisions are compared to data at RHIC energies, and
discussed for several extrapolations to LHC energies. 
\\[.4cm]
\newpage

\section{Introduction}
Stopping and particle production in relativistic heavy-ion collisions
at the highest energies available at RHIC and LHC offer sensitive 
tools to test the nonequilibrium-statistical properties of these 
systems. Analytically soluble nonequilibrium-statistical models 
\cite{wol99} not 
only allow to accurately describe a fairly large amount of phenomena,
but also to predict results across the gap in 
center-of-mass energy from the highest RHIC energy in Au + Au 
collisions of $\sqrt{s_{NN}}$=200 GeV \cite{bea04,bac03} to the LHC energy of 
$\sqrt{s_{NN}}$= 5.52 TeV in Pb + Pb collisions. 

Net-baryon (more precisely, net-proton) distributions have 
proven to be sensitive indicators for local equilibration,
collective expansion, and 
deconfinement in heavy relativistic systems \cite{wol03}.
This is reconsidered in the analysis of SPS- and RHIC-results
within the Relativistic Diffusion Model (RDM) for three sources in Sect. 2.
In this work, we use the dependence of the transport coefficients on the 
center-of-mass energy from AGS via SPS to RHIC for  
predictions of net-baryon rapidity distributions at LHC, Sect. 3.

The model underlines the nonequilibrium-statistical
features of relativistic heavy-ion collisions, but it also encompasses
kinetic (thermal) equilibrium of the system for times that are
sufficiently larger than the relaxation times of the relevant
variables.

It is of particular interest in relativistic collisions of
heavy systems to determine the fraction of produced particles
that attains - or comes very close to - local statistical equilibrium.
In the three-sources RDM, these are the particles produced in
the midrapidity source. Hence we analyze Au + Au and Cu + Cu 
pseudorapidity distributions of produced particles
at RHIC energies from $\sqrt{s_{NN}}$=19.6 - 200 GeV.
We determine the transport coefficients and numbers of
produced particles as functions
of the incident energies in Sect. 4. 

For several reasonably motivated
extrapolations of the transport coefficients we then
calculate and discuss a range of resulting distribution functions for
produced charged hadrons in Pb + Pb collisions 
at LHC energies. The conclusions are drawn in Sect. 5.
 
\section{Net-baryon rapidity spectra}
In the Relativistic Diffusion Model, the
net-baryon rapidity distribution at RHIC energies
emerges from a superposition of the beam-like nonequilibrium
components that are broadened in rapidity space through
diffusion due to soft (hadronic, low $p_{\perp}$) collisions and
particle creations, and a
near-equilibrium (thermal) component at midrapidity
that arises - among other processes -
from hard (partonic, high $p_{\perp}$) processes,
and may indicate local quark-gluon plasma (QGP) formation.

The time evolution of the distribution
functions is governed by a Fokker-Planck
equation (FPE) in rapidity space
\cite{wol99,wol03,alb00,ryb02,biy02} 
%In the more general 
%case of
%nonextensive (non-additive) statistics \cite{tsa88} that accounts
%for long-range interactions and violations
%of Boltzmann's Sto{\ss}zahlansatz \cite{alb00} as well as for
%non-Markovian memory (strong coupling) effects \cite{wol04},
%the resulting FPE for the rapidity distribution function
%R(y,t) in the center-of-mass frame is
%\cite{wol03}

\begin{equation}
\frac{\partial}{\partial t}[ R(y,t)]^{\mu}=-\frac{\partial}
{\partial y}\Bigl[J(y)[R(y,t)]^{\mu}\Bigr]+D_{y}
\frac{\partial^2}{\partial y^2}[R(y,t)]^{\nu} 
\label{fpenl}
\end{equation}\\
with the rapidity $y=0.5\cdot \ln((E+p)/(E-p))$.
The rapidity diffusion coefficient $D_{y}$ that contains the
microscopic physics accounts for the broadening of the
rapidity distributions due to interactions and particle creations, and
it is related to the drift term $J(y)$ by means of a 
dissipation-fluctuation theorem (Einstein relation) which is used to actually 
calculate $D_{y}$ in the weak-coupling limit \cite{wol99}.
The drift $J(y)$ determines the shift of the mean rapidities
towards the central value, and linear and nonlinear 
forms have been discussed.

Here we use $\mu$ = 1 (due to norm conservation)
and $q = 2 - \nu$ = 1 corresponding to the standard FPE,
and a linear drift function
 \begin{equation}
J(y)=(y_{eq}- y)/\tau_{y}
\label{dri}
\end{equation}
with the rapidity relaxation time $\tau_{y}$, and the equilibrium 
value Ê$y_{eq}$ of the rapidity \cite{wol99,wols06}. This is
the so-called Uhlenbeck-Ornstein \cite{uhl30} process, applied to the
relativistic invariant rapidity for the three components  
$R_{k}(y,t)$ ($k$=1,2,3) of the distribution function
in rapidity space
\cite{wol99,wol03,biy02} 

\begin{equation}
\frac{\partial}{\partial t}R_{k}(y,t)=
\frac{1}{\tau_{y}}\frac{\partial}
{\partial y}\Bigl[(y-y_{eq})\cdot R_{k}(y,t)\Bigr]
+D_{y}^{k}\frac{\partial^2}{\partial^{2} y}
 R_{k}(y,t).
\label{fpe}
\end{equation}\\

Since the equation is linear, a superposition of the distribution
functions \cite{wol99,wol03} emerging from $R_{1,2}(y,t=0)=\delta(y\mp y_{b})$
and $R_{3}(y,t=0)=\delta(y-y_{eq})$  
with mean values and variances obtained analytically from the moments 
equations yields the exact solution. With two sources and
 $D_{y}$ calculated from the 
dissipation-fluctuation theorem, net-baryon rapidity spectra at
low SIS-energies (about 1 GeV per particle) are well reproduced,
whereas at AGS and SPS energies one finds discrepancies 
\cite{wol99} to the data that rise strongly with $\sqrt{s}$ due to
the collective longitudinal expansion of the system.

With an effective 
diffusion coefficient $D_{y}^{eff}$ that includes not only
the random behaviour of the particles, but also the effect of expansion,
however, the available net-proton data for central Au+Au collisions
at AGS, and Pb+Pb at SPS \cite{app99} can be reproduced precisely
with only two sources in the RDM. The
corresponding rapidity width coefficient
\begin{equation}
\Gamma_{y}^{eff}=[8\cdot \ln(2) \cdot D_{y}^{eff}\cdot \tau_{y}]^{1/2}
\label{gam}
\end{equation}
is shown in Fig.1 (middle frame) as function of the center-of-mass energy. 
The quotient of interaction and 
relaxation time
$\tau_{{int}}/\tau_{y}$ (Fig.1, upper frame) is determined from the 
peak positions of the data, and the corresponding longitudinal 
expansion velocity (Fig.1, lower frame) is obtained as in \cite{wol06}.
The rapidity distributions are shown in Fig. 2.

Extending the model to RHIC energies of 200 GeV, one finds \cite{wol03}
that within 
the linear approach with $q=1$ and two sources, it is impossible to
reproduce the BRAHMS net-proton data \cite{bea04} due to the high 
midrapidity yield.

This has recently been confirmed in an independent calculation by 
Alberico et al. \cite{al06} that uses the nonlinear approach,
but also investigates the linear case.
It has therefore been proposed in \cite{wol03} that an
expanding midrapidity source emerges. With this conjecture, 
the RHIC data can be
interpreted rather precisely in the linear $q=1$ framework.
A fraction of
$Z_{eq}\simeq 22$ net protons (55 net baryons) near midrapidity
reaches local statistical equilibrium
in the longitudinal degrees of freedom, Fig. 2. The variance of
the equilibrium distribution $R_{eq}(y)$ at midrapidity is broadened
as compared to the Boltzmann result 
due to collective (multiparticle) effects.
This corresponds to a longitudinal expansion (longitudinal flow)
velocity of the locally equilibrated subsystem
as in hydrodynamical descriptions. Here we obtain the expansion
velocity as proposed in \cite{wol06} 

\begin{equation}
v_{coll}^{||}=\left[1-\left[\frac{m_{p}}{m_{p}+\frac{1}{2}\cdot(T_{eff}-T)}
\right]^{2}\right]^{1/2}
\label{vco}
\end{equation}
with the limiting cases $v_{coll}^{||}=1$ for $T_{eff}>>T$, and
$v_{coll}^{||}=0$ for $T_{eff}=T$. The kinetic freezeout 
temperature is T$\simeq$ 110 MeV at RHIC energies, and $T_{eff}$ is
the effective temperature required to obtain the correct width of the
midrapidity plateau. The proton mass is $m_{p}$.

The enhanced width of the midrapidity valley may also
be accounted for in a non-extensive framework with
two sources, through a value of $q=1.485$ \cite{al06,tsa88}. 
We believe, however, that a value of $q$ larger than one
covers the physical effect of longitudinal expansion
which can be determined explicitly 
based on ordinary Boltzmann statistics.

Microscopically, the baryon transport over 4-5
units of rapidity to the
equilibrated midrapidity region requires
processes such as the nonperturbative
gluon junction mechanism \cite{ros80} 
to produce the observed central valley. 

Macroscopically, the complete solution of (\ref{fpe}) 
in the $q=1$ case is a linear superposition
of nonequilibrium and local equilibrium distributions. 
The net-baryon rapidity 
distribution becomes

\begin{equation}
\frac{dN(y,t=\tau_{int})}{dy}=N_{1}R_{1}(y,\tau_{int})+N_{2}R_{2}(y,\tau_{int})
+N_{eq}R_{eq}^{loc}(y)
\label{normloc}
\end{equation}\\
with the interaction time $\tau_{int}$ (total integration time of the
differential equation). The number of net baryons 
(here: net protons) in local equilibrium
near midrapidity is $N_{eq}$, and $N_{1} + N_{2} + N_{eq}$
is equal to the total 
number of net baryons (corresponding to 158 net protons for central Au+Au).
This yields a good representation of the 
BRAHMS data \cite{bea04}, as was already emphasized in \cite{wol06}. 
Here the integration is stopped at the value of $\tau_{int}/\tau_{y}$
that yields the best agreement with the data.
\section{From RHIC to LHC energies} 
Based on the results for net protons from AGS to RHIC energies,
we have extrapolated the diffusion-model parameters as functions of
the center-of-mass energy to LHC energies, Fig. 1. 
To obtain the net-proton distributions at LHC energies of
$\sqrt{s_{NN}}=14Z/A$ TeV $\simeq 5.52$ TeV we use the following 
extrapolations for the time parameter and the effective width
coefficient
with the logarithm of $\sqrt{s_{NN}}$:
\begin{equation}
\frac{\tau_{int}}{\tau_{y}}(\sqrt{s_{NN}})=
2exp(-1.1\log(\sqrt{s_{NN}}))+0.11
\label{textr}
\end{equation}
\begin{equation}
\Gamma_{y}^{eff}(\sqrt{s_{NN}})=1.8á\log(\sqrt{s_{NN}}) + 0.2.
\label{gaextr}
\end{equation} 

This allows to obtain a first approximation of
the net-proton distribution at LHC energies as shown in Fig. 2.
Here, the solid curve is for a midrapidity source with a particle
content of 14{\%} as at the highest RHIC energy,
whereas the dashed curve corresponds to
a particle content of only 7{\%}.

Due to the small value of $\tau_{int}/\tau_{y}=0.14$ that is obtained
from the extrapolation, the distribution functions at LHC energies 
extend well beyond the initial beam rapidity shown by an arrow.
Hence, this result will have to be corrected at large
rapidities since it is likely to violate kinematic constraints.

It should be mentioned that the dependence of the time parameter 
$\tau_{int}/\tau_{y}$ on $\sqrt{s_{NN}}$ can not be expected to be 
purely exponential over the whole energy scale. In particular, the
curve is known to level off at low energies, as is evident from an 
earlier investigation of symmetric systems at SIS energies around 
$\sqrt{s_{NN}}\simeq$ 2 GeV \cite{wol99}.
Such a behaviour is plausible from the
relation between the drift in rapidity space, and the energy loss 
that is associated with it. As has been pointed out by Koch 
\cite{ko06}, the mean rapidity loss per step can be written as
\begin{equation}
<\Delta y> 
\propto \frac{yN_{part}}{\sinh(y)} \propto \frac{\tau_{int}}{\tau_{y}} .
\label{dely}
\end{equation} 
This result is based on the usual relation between energy and 
rapidity
\begin{equation}
E=m\cosh(y) ,
\label{Ey}
\end{equation}
or differentially
\begin{equation}
dE=m\sinh(y)dy .
\label{dEy}
\end{equation}
On the other hand, the differential energy loss is proportional 
to the particle mass, the number of participants, and the
rapidity
\begin{equation}
dE \propto myN_{part}
\label{dENy}
\end{equation}
such that the above Eq.(\ref{dely}) results. The expression is shown 
as a dashed curve in the upper frame of Fig.1. It levels off 
at small energies below 2 GeV ($\tau_{int}/\tau_{y}=1.98$
at 1 GeV) as a consequence of the relation between
energy- and rapidity loss for net baryons. At large energies, the
result of (\ref{dely}) shows a steeper fall-off which is 
difficult to reconcile with the data at the highest RHIC energy.
For the extrapolation to LHC energies, we therefore use the 
exponential form in the net-proton case.

The fall of $\tau_{int}/\tau_{y}$ with rising $\sqrt{s_{NN}}$ does
not imply that the lifetime of the locally equilibrated region becomes
shorter with rising energy. According to common belief, the QGP-lifetime is
expected to increase as the energy rises from RHIC to LHC.

\section{Produced charged hadrons in the RDM}
The Relativistic Diffusion Model \cite{wol99} in its linear form with explicit 
treatment of the collective expansion \cite{wol03}, or in its nonlinear version
with implicit consideration of the collective effects, is also 
suitable for the description and prediction of rapidity 
distributions of produced charged hadrons. Although the model results 
are somewhat more ambiguous because the initial conditions are not 
sharply defined as in case of the net-proton distributions, we 
proceed with the calculation of distribution functions for
produced charged hadrons at RHIC and LHC energies.

If particle identification is not available, one has to
convert the results to pseudorapidity space, 
$\eta=-$ln[tan($\theta / 2)]$ with the scattering angle $\theta$.
The conversion from $y-$ to $\eta-$
space of the rapidity density
\begin{equation}
\frac{dN}{d\eta}=\frac{dN}{dy}\frac{dy}{d\eta}=\frac{p}{E}\frac{dN}{dy}=
J(\eta,\langle m\rangle/\langle p_{T}\rangle)\frac{dN}{dy} 
\label{deta}
\end{equation}
is performed through the Jacobian
\begin{equation}
J(\eta,\langle m\rangle/\langle p_{T}\rangle) 
 = \cosh({\eta})\cdot [1+(\langle m\rangle/\langle p_{T}\rangle)^{2}
+\sinh^{2}(\eta)]^{-1/2}.
\label{jac}
\end{equation}
We approximate the average mass $<m>$ of produced charged hadrons in the
central region by the pion mass $m_{\pi}$, and use a
mean transverse momentum $<p_{T}>$ = 0.4 GeV/c.

In the linear two-sources version, the RDM had been applied to 
pseudorapidity distributions of produced charged hadrons in Au+Au 
collisions at RHIC energies of 130 GeV and 200 GeV 
by Biyajima et al. \cite{biy02}. Although the results were 
satisfactory, it soon turned out from the above net-proton results
\cite{wol03},
and from general considerations, that an additional midrapidity 
source is required \cite{biya04,wols06}. 

Asymmetric relativistic systems are
particularly sensitive to the details of the diffusion-model 
calculation, as was shown recently in our description of the d + Au 
system at 200 GeV in the three-sources model \cite{wols06}. Here, an 
accurate modelling of the gradual approach of the system to thermal 
equilibrium was obtained. In particular, the dependence of the pseudorapidity 
distribution functions on centrality was precisely described.
In the present investigation, however, we concentrate on
symmetric systems because these will be of main interest at LHC.
Our focus is on central collisions.

To allow for an extrapolation of our results 
for symmetric systems to LHC energies, we 
first perform RDM-calculations for Cu + Cu and Au + Au 
collisions at RHIC energies 
from 19.6 GeV via 62.4 GeV, 130 GeV to 200 GeV \cite{kui06}.

Typical results for 
central collisions of Au + Au at three energies are shown in Fig. 3  
compared with PHOBOS data \cite{bac03}, and  of Cu + Cu at
two energies in Fig. 4 compared with preliminary PHOBOS data 
\cite{nou05}. At the lowest energy, only 
two sources are needed for the optimization of the RDM-parameters in 
a $\chi^{2}$-fit, whereas three sources are indeed required at the higher 
energies. The particle number in the midrapidity source is
indicated in the figures. At the highest energy of 200 GeV,
the Cu + Cu system requires a smaller percentage of particles in the
midrapidity source compared to Au + Au. This is consistent with the
assumption that heavier systems are more likely to produce a locally
equilibrated quark-gluon plasma.

The $\chi^{2}$-minimization program has been written in Mathematica 
for the purpose of this work \cite{kui06}. In a previous 
investigation \cite{wols06}, the CERN minuit code \cite{jam81} was 
used. We have verified that for 200 GeV Au + Au, the results are 
identical. When the execution stops at minimum $\chi^{2}$, the
values of the time parameter, and of the effective widths of
the partial distribution functions are determined.

The parameters of these calculations are summarized in Table 1. The
number of particles produced in the three sources, the time parameters 
and the effective widths of the partial distributions (including the
time evolution) are shown 
together with $\chi^{2}/d.o.f.$. The number of degrees of 
freedom ($d.o.f.$) is the number of data points minus the number
of free parameters.

For produced charged hadrons, the results for the time parameter
$\tau_{int}/\tau_{y}$ as function of energy (Fig. 5, upper frame)
are found to be
significantly larger than the corresponding values for net protons 
(Fig.1).
We have extrapolated the time parameter to
LHC energies. The upper curve in Fig. 5 has the functional dependence 
on energy that we have discussed for net baryons
\begin{equation}
\frac{\tau_{int}}{\tau_{y}} 
\propto \frac{yN_{part}}{\sinh(y)} 
\label{hdely}
\end{equation} 
whereas the lower curve assumes an exponential dependence.

The resulting partial widths as functions of energy within the RHIC 
range for Au + Au are shown in the 
middle frames of Fig. 5 for both peripheral and midrapidity
sources, which are not assumed to be equal for produced hadrons.
As for net protons, the widths are effective values:
they include the effect of collective expansion. We use 
log-extrapolations of the widths to LHC energies which
yield the values given in the figures. Here we have plotted the
values resulting from the $\chi^{2}$-minimization that include the
time evolution up to $\tau_{int}$ 
\begin{equation}
\Gamma_{1,2,eq}^{eff}=[8 \ln(2) \cdot D_{1,2,eq}^{eff}\cdot \tau_{y}
\cdot (1-\exp(-2\tau_{int}/\tau_{y}))]^{1/2}
\label{gamh}
\end{equation}
The total number of produced charged particles relative to 
the number of participants resulting from the $\chi^{2}$-minimizations
in various systems that we have investigated in the course
of this work (including the asymmetric d + Au case)
is shown in Fig. 6 as function of the 
center-of-mass energy. Extrapolating to LHC 
energies, we obtain 26.5 produced charged hadrons per participant
pair in 
central collisions (0-6{\%}). With an average number of 359.3 
participants in the most central bin for Pb + Pb, this yields a
total of 9520 produced charged hadrons. 

%Using this estimate together 
%with a central source containing 50{\%} of the produced charged 
%hadrons, and the extrapolation of the diffusion-model parameters for
%produced particles to LHC from Fig. 4, we obtain the result shown in 
%Fig. 6 (A) for 
%the pseudorapidity distribution of charged hadrons in central Pb + Pb 
%collisions at LHC energies. For different extrapolations of the 
%rapidity density at midrapidity, 
%substantially different results are obtained. As examples, (B) shows
%a logarithmic extrapolation of the midrapidity yield as performed in
%\cite{arm05} together with an increase by 30{\%} of the width, 
%whereas (C) is the distribution function based on the midrapidity 
%result in the so-called saturation model \cite{arm05}
%together with an increase by 60{\%} of the width.

In the Jacobian (\ref{jac}), the mean transverse momentum $<p_{T}>$
becomes significantly larger at LHC energies. Due to the increasing 
number of produced hadrons, the mean mass - which tends to approach 
the pion mass - decreases,  such that $(\langle m\rangle/\langle 
p_{T}\rangle)^{2}<<1$ with increasing $\sqrt{s_{NN}}$, and the Jacobian
\begin{equation}
\frac{dy}{d\eta}=\frac{\cosh(\eta)}{\sqrt{1+\frac{<m>^2}
{<p_t>^2}+\sinh^2(\eta)}}\approx\frac{\cosh(\eta)}{\sqrt{1+\sinh^2(\eta)}}=1
\end{equation}
becomes very close to one at sufficiently high energy. Hence we use 
$dy/d\eta = 1$
in our predictions of pseudorapidity distributions at LHC energies.

In view of the uncertainties that arise from the large energy gap
between RHIC and LHC, we have tested variations of the 
diffusion-model parameters that indicate the spreading
of our diffusion-model results within reasonable limits.
Based on the extra\-polations of 
the time parameter and the effective widths to LHC energies
shown in Fig. 5 and a mean particle content of the midrapidity
source of 50{\%}, we display in Fig. 7 calculated 
pseudorapidity distribution functions 
for charged particles at LHC energies for successive variations
of the transport parameters.

In the upper (first) frame results  
for three different values of the time parameter
x=$\tau_{int}/\tau_{y}$ are shown. In the second frame the
dependence of the distribution function on the widths of the
peripheral distributions is displayed. The third frame
gives the dependence on the
width of the midrapidity source, and the fourth frame the
dependence on the particle content of the midrapidity source.
In each case, the values of those parameters that are not
varied are taken from the extrapolation in Fig. 5 with 
the time parameter from (\ref{hdely}). Considerables changes in
the shapes of the distribution functions are observed, in particular,
for varying particle content of the central source.
It is presently not possible to predict the particle content
of the midrapidity source at LHC energies on the basis of
the two Au + Au results at 130 GeV and 200 GeV.

Our diffusion-model result with extrapolated parameters from Fig. 5
and 50{\%} particle content in the central source is shown in
Fig. 8, curve $[A]$. For comparison, the RHIC result
for 
$\sqrt{s_{NN}}                                                                          $=200 GeV of Fig. 3 is redisplayed in the upper
frame together with this prediction. It is significantly narrower
than the LHC result. Here the time parameter is
taken from (\ref{hdely}). Curve $[B]$ in the lower frame
is obtained with the exponential extrapolation of the time parameter
as shown in Fig. 5. This yields a smaller value of $x$ and hence,
a broader distribution function. Values of the particle content of the
midrapidity source, and of the widths are given in the caption.

We have also obtained diffusion-model results using the extrapolations of
the number of produced particles at midrapidity given by other
authors. In particular, curve $[C]$ in Fig. 8 uses a logarithmic
extrapolation of the midrapidity value that yields dN/d$\eta \simeq$
1100 \cite{arm05}, whereas
the saturation model \cite{arm05} predicts dN/d$\eta \simeq$ 1800 at
midrapidity, with the resulting diffusion-model distribution $[D]$. 
The forthcoming LHC data are likely to be in between the limiting 
cases $[A]$ and $[D]$. 

Whereas the precise midrapidity value will have to
be determined from the data, it is the detailed shape of the
distribution function that will be of interest
in order to determine to what extent the
system reaches statistical equilibrium at LHC energies.

It is interesting to compare our results with predictions of other
models that are not based on nonequilibrium-statistical
mechanics, but instead on QCD.
In particular, there are calculations within the framework of 
the Parton Saturation Model that not only predict the midrapidity 
value, but also the full distribution function at RHIC energies, and
also at LHC \cite{nar05}. These calculations are based on a classical 
effective theory that describes the gluon distribution in large nuclei 
at high energies where saturation might occur at a critical momentum 
scale, to form a so-called Color Glass Condensate (CGC) \cite{mcl94}.

Although this assumption has a clear and reasonable physical 
basis, the resulting pseudorapidity 
distribution functions for produced charged hadrons 
at the available RHIC energies do not appear to match the precision of our 
phenomenological three-sources
diffusion-model results when compared to the existing data. At LHC
energies, the overall CGC-distribution for a given midrapidity value
as obtained with the assumption of a constant $\alpha_{s}$ for strong 
coupling is slightly narrower than the corresponding diffusion-model result.

Additional consideration of a running coupling gives a midrapidity value that
is of the order of 10{\%} smaller; another uncertainty arises from the 
extrapolation of the saturation scale to LHC 
energies. Various predictions for central rapidity densities and
pseudorapidity distributions at RHIC and LHC energies had been
summarized e.g. in \cite{arm00}, where also the differences among the existing
models - including hydrodynamical and pQCD approaches and 
their numerical implementations - had been discussed. It appears, however, that
the analytical diffusion-model approach provides better results
when compared in detail to the experimental distribution functions
at RHIC energies.

\newpage
\section{Conclusion}
 We have described net-proton and charged-hadron distributions
in collisions of heavy systems at  SPS and RHIC energies in
a Relativistic Diffusion Model (RDM) for
multiparticle interactions.
Analytical results for the rapidity
distribution of net protons in central collisions are found to be in 
good agreement with the available data. An extrapolation of the
rapidity distributions for net protons to LHC energies has been 
performed. The precise number of particles in the midrapidity source 
remains uncertain at LHC energies and will have to be determined from experiment.

At RHIC - and most likely at LHC -, a significant
fraction of the net protons (about 14 per cent at 200 GeV for Au + Au) 
reaches local statistical equilibrium
in a fast and discontinuous transition which we believe to
indicate parton deconfinement. 
%The corresponding number at LHC will
%have to be determined from the data.
%A detailed investigation of
%the flat midrapidity valley found at RHIC, and of its 
%dependence towards LHC energies is very 
%promising. 

In the three-sources RDM, we have calculated
pseudorapidity distributions of produced charged hadrons. 
The diffusion-model parameters have been optimized
in a $\chi^{2}-$minimization with respect to the available
PHOBOS data in Au+Au and Cu+Cu \cite{kui06} at RHIC.
Excellent results for the energy dependence of the
distribution functions have 
been obtained.
%,although they remain more ambiguous as compared to the 
%net-proton case. 

In an extrapolation to LHC energies, the 
pseudorapidity distribution for produced charged hadrons has been 
calculated. Here the essential parameters relaxation time, diffusion
coefficients or widths of the distribution
functions of the three sources, and number of particles 
in the local equilibrium source 
will have to be adjusted once the ALICE data for Pb + Pb have become 
available in 2009.

\newpage
\rm
%\small

\newpage
\rm
{\bf Table 1.} Produced charged hadrons for central (0-6{\%}) collisions
of Cu + Cu at RHIC energies of
$\sqrt{s_{NN}}$ = 62.4 GeV and 200 GeV, of Au + Au at 
 19.6 GeV, 130 GeV, and 200 GeV, and of Pb + Pb at LHC energies
 in the Relativistic Diffusion Model.
 The number of produced charged particles is $N_{ch}^{tot}$ with 
 $N_{ch}^{1,2}$ for the sources 1 and 2 and $N_{ch}^{eq}$ for the equilibrium
source, the percentage of
charged particles produced in the locally thermalized source is 
$n_{ch}^{eq}$.
The ratio $\tau_{int}/\tau_{{y}}$ determines how fast the system 
of produced charged particles equilibrates in rapidity space.
The effective widths of the peripheral sources are
$\Gamma_{1,2}^{eff}$, of the midrapidity source $\Gamma_{eq}^{eff}$. 
The $\chi^{2}/d.o.f.$ is as shown in Figs. 3, 4, with $d.o.f.=$number 
of data points - number of free parameters. 
%The constraint is  
%discussed in the text. The result of a fit without constraints as in
%\cite{wols06} is $\chi^{2}_{min}/d.o.f..$
\\[1.5cm]
	\renewcommand{\arraystretch}{1.5}
		\begin{table}[h]	
			\begin{center}
			\begin{tabular}{|l||r|r|r||r|r|r|r||r|r|r||r|r|}
				\hline
				\rule[-3mm]{0mm}{8mm}	
				system and energy 		&
				$N_{ch}^{tot}$ 			&
				$N_{ch}^{1,2}$ 		&
				$N_{ch}^{eq}$ 			&
				$n_{ch}^{eq}(\%)$ 		&
				$\frac{\tau_{int}}{\tau_y}$ 	&
				$\Gamma_{1,2}^{eff}$ 	&
				$\Gamma_{eq}^{eff}$ 	&
				$\frac{\chi^2}{d.o.f.}$	\\	
			%	$\frac{\chi_{min}^2}{d.o.f.}$\\
				\hline\hline
				
				Cu+Cu 62.4 GeV		& 
				825					& 
				400					& 
				26					& 
				3.2					& 
				1.12					& 
				3.70					& 
				5.15					& 
				$\frac{4.7}{49}$  \\			
			%	$\frac{1.9}{49}$	\\
				
				\hline
				Cu+Cu 200 GeV		& 
				1474					& 
				685					& 
				105					& 
				7.1					& 
				1.08					& 
				4.03					& 
				2.45					& 
				$\frac{2.0}{49}$  \\		
			%	$\frac{0.9}{49}$	\\
				
				\hline			
				Au+Au 19.6 GeV		& 
				1692					& 
				846					& 
				-					& 
				-					& 
				1.23					& 
				2.90& -				& 
				$\frac{0.7}{28}$  \\		
				%$\frac{0.3}{26}$	\\ 
				 
				\hline 
				Au+Au 130 GeV		& 
				4233					& 
				1837					& 
				560					& 
				13.2					& 
				1.02					& 
				3.56					& 
				2.64					& 
				$\frac{3.7}{49}$  \\		
			%	$\frac{0.8}{49}$	\\
				
				\hline 
				Au+Au 200 GeV		& 
				5123					& 
				1887					& 
				1349					& 
				26.3					& 
				0.93					& 
				3.51					& 
				3.20					& 
				$\frac{1.1}{49}$ \\		
				%$\frac{0.7}{49}$	\\  
				\hline
					Pb+Pb 5520 GeV		& 
				9520					& 
				2380					& 
				4760					& 
				50					& 
				0.91					& 
				4.6					& 
				7.5					& 
				- \\		
				%$\frac{0.7}{49}$	\\  
				\hline
			\end{tabular}
			\end{center}
		%\caption{}
		\end{table}
	\renewcommand{\arraystretch}{1}
	
\newpage
\bf
Figure captions\\\\
\rm
\begin{description}
\item[Fig. 1.]
Dependence of the Diffusion-Model parameters for heavy systems
(Au + Au at AGS and RHIC, Pb + Pb at SPS and LHC) on the center-of-mass 
energy $\sqrt{s_{NN}}$, with extrapolations
to LHC (top to bottom frame): Quotient of interaction time and relaxation 
time - see text for dashed curve; rapidity 
width coefficient including collective expansion; and longitudinal 
collective velocity. The results are for net-proton rapidity
distributions.
\item[FIG. 2.]
Net-proton rapidity spectra in the Relativistic Diffusion Model
(RDM), solid curves: Transition from the double-humped shape at SPS 
energies of $\sqrt{s_{NN}}$ = 17.3 GeV to a broad midrapidity valley
in the three-sources model at RHIC (200 GeV) and LHC (5.52 TeV). 
The two curves at LHC energies contain 14{\%} and 7{\%} of the 
net protons in the
midrapidity source, respectively. Kinematic constraints will modify 
the LHC-distributions at large values of $y$. 
Data are from NA49 at SPS \cite{app99}, 
and BRAHMS at RHIC \cite{bea04}.
\item[FIG. 3.]
Calculated pseudorapidity distributions of produced charged hadrons from
central Au + Au collisions at $\sqrt{s_{NN}}$ = 19.6 GeV, 130 GeV and
200 GeV. The analytical 
RDM-solutions (two sources at the lowest, three at the higher 
energies) are optimized in a $\chi^{2}$-fit with respect to 
PHOBOS data \cite{bac03}. The percentage of particles in the midrapidity source 
is indicated.
\item[FIG. 4.]
Calculated pseudorapidity distributions of produced charged hadrons from
central Cu + Cu collisions at $\sqrt{s_{NN}}$ = 62.4 GeV and 200 GeV.
The analytical 
RDM-solutions for three sources are optimized in a $\chi^{2}$-fit 
with respect to PHOBOS data \cite{nou05}. The percentage of particles in 
the midrapidity source is indicated.
\item[Fig. 5.]
Dependence of the Diffusion-Model parameters for heavy systems
(Au + Au at AGS and RHIC, Pb + Pb at LHC) on the center-of-mass 
energy $\sqrt{s_{NN}}$, with extrapolations
to LHC (arrows):  Quotient of interaction time and relaxation 
time for sinh- and exponential extrapolation (upper frame); rapidity 
width of the peripheral sources including collective expansion (middle
frame); width of the midrapidity source (lower frame).
The results are for charged-hadron rapidity distributions.
\item[FIG. 6.]
Total number of produced charged hadrons per number of participant
nucleons as function of the center-of-mass energy as obtained from
the RDM-results for d+Au, Cu+Cu, Au+Au, and for Pb+Pb \cite{app99} 
pseudorapidity 
distributions. The extrapolation to LHC energies used in this work is 
also shown.
%\item[FIG. 6.]
%Pseudorapidity distributions of produced charged hadrons 
%for central Pb + Pb collisions at LHC energies of 5520 GeV with (A)
%a charged-particle content in the central source of 50{\%},
%extrapolated transport parameters from Fig. 4, and total number of 
%produced charged hadrons from Fig. 5.
%(B) shows a corresponding calculation with 
%a log-extrapolation \cite{arm05} of the midrapidity production rate
%combined with an increase of the widths by 30{\%}. (C) uses the
%saturation-model prediction of the midrapidity rate \cite{arm05}
%combined with a 60{\%} increase of the widths.
\item[FIG. 7.]
Pseudorapidity distributions of produced charged hadrons 
for central Pb + Pb collisions at LHC energies of 5520 GeV with
a charged-particle content in the central source of 50{\%},
extrapolated transport parameters from Fig. 5, and total number of 
produced charged hadrons from Fig. 6. Upper (first) frame:
Results for three different values of the time parameter
$x=\tau_{int}/\tau_{y}$. Second frame: 
Dependence of the distribution function on the widths of the
peripheral distributions. The result for $\Gamma_{1,2}$=4.6
is for the extrapolated values of all parameters
as in curve $[A]$ of Fig. 8.
Third frame:
Dependence on the
width of the midrapidity source. Fourth frame: 
Dependence on the particle content of the midrapidity source.
The values of those parameters that are not
varied are taken from the extrapolation in Fig. 5. 
\item[FIG. 8.]
Calculated pseudorapidity distributions of produced charged hadrons 
for central Au + Au collisions at RHIC compared with 200 A GeV
PHOBOS data \cite{bac03}, and diffusion-model 
extrapolation to Pb + Pb at LHC energies of 5520 GeV with
a charged-particle content in the central source of 50{\%},
transport parameters from Fig. 5, and total number of 
produced charged hadrons from Fig. 6 (upper frame).
Lower Frame: Curve $[A]$ is the diffusion-model result as in the
upper frame.
Curve $[B]$ 
is obtained with the exponential extrapolation of the time parameter
shown in Fig. 5 and $\Gamma_{1,2}=5.9$ (cf. Fig. 7); here the
particle content of the midrapidity source is $40\%$.
Curve $[C]$ uses a logarithmic
extrapolation of the midrapidity value 
that yields 1100 \cite{arm05}, with $\Gamma_{1,2}=5.9$.
The saturation model \cite{arm05} predicts $dN/d\eta \simeq$ 1800 at
midrapidty with the resulting distribution $[D]$ using 
$\Gamma_{1,2}=5.9$.

\end{description}
\newpage
\vspace{1cm}
\includegraphics[width=12cm]{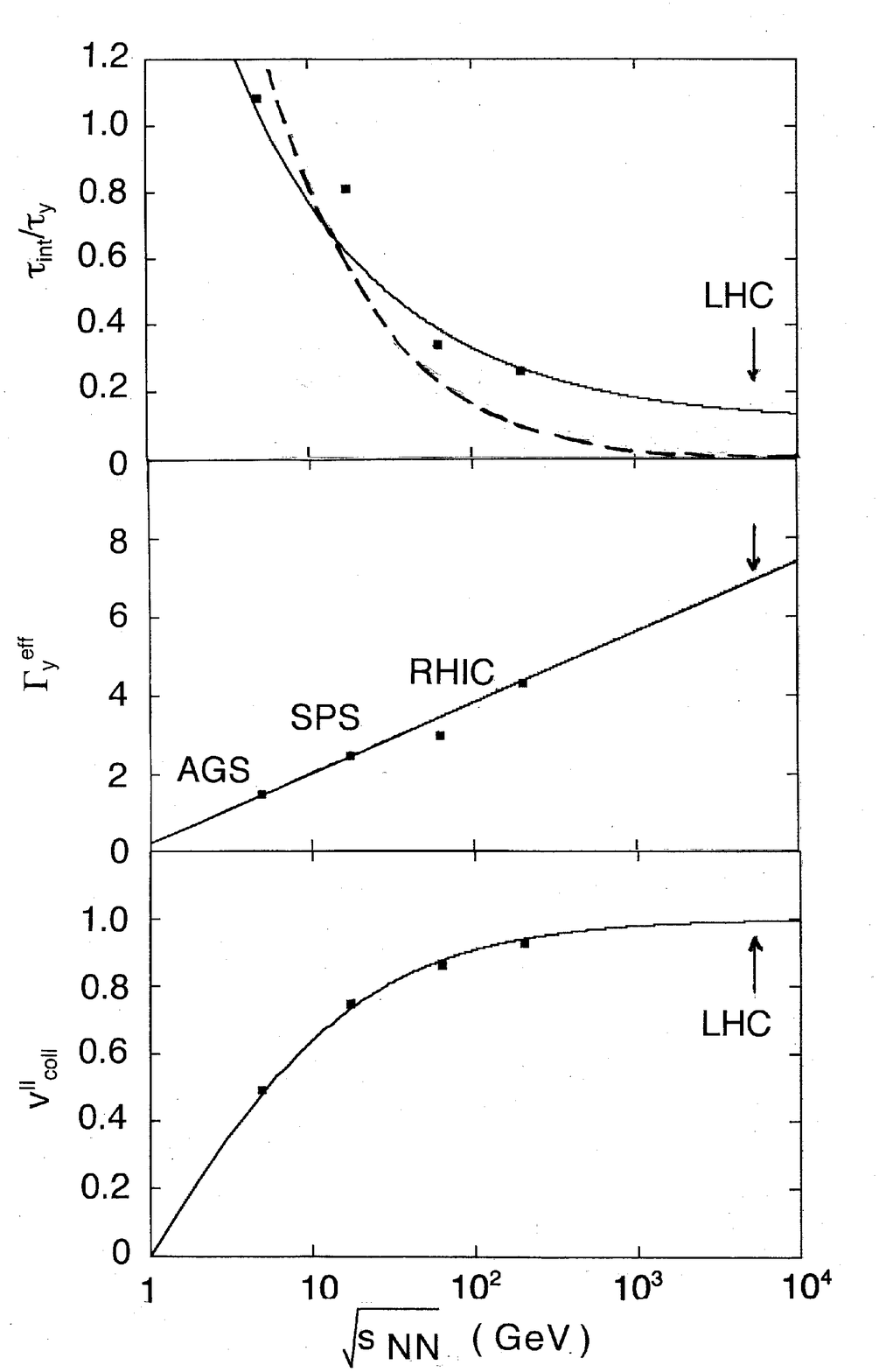}
\newpage
\vspace{1cm}
\includegraphics[width=15.2cm]{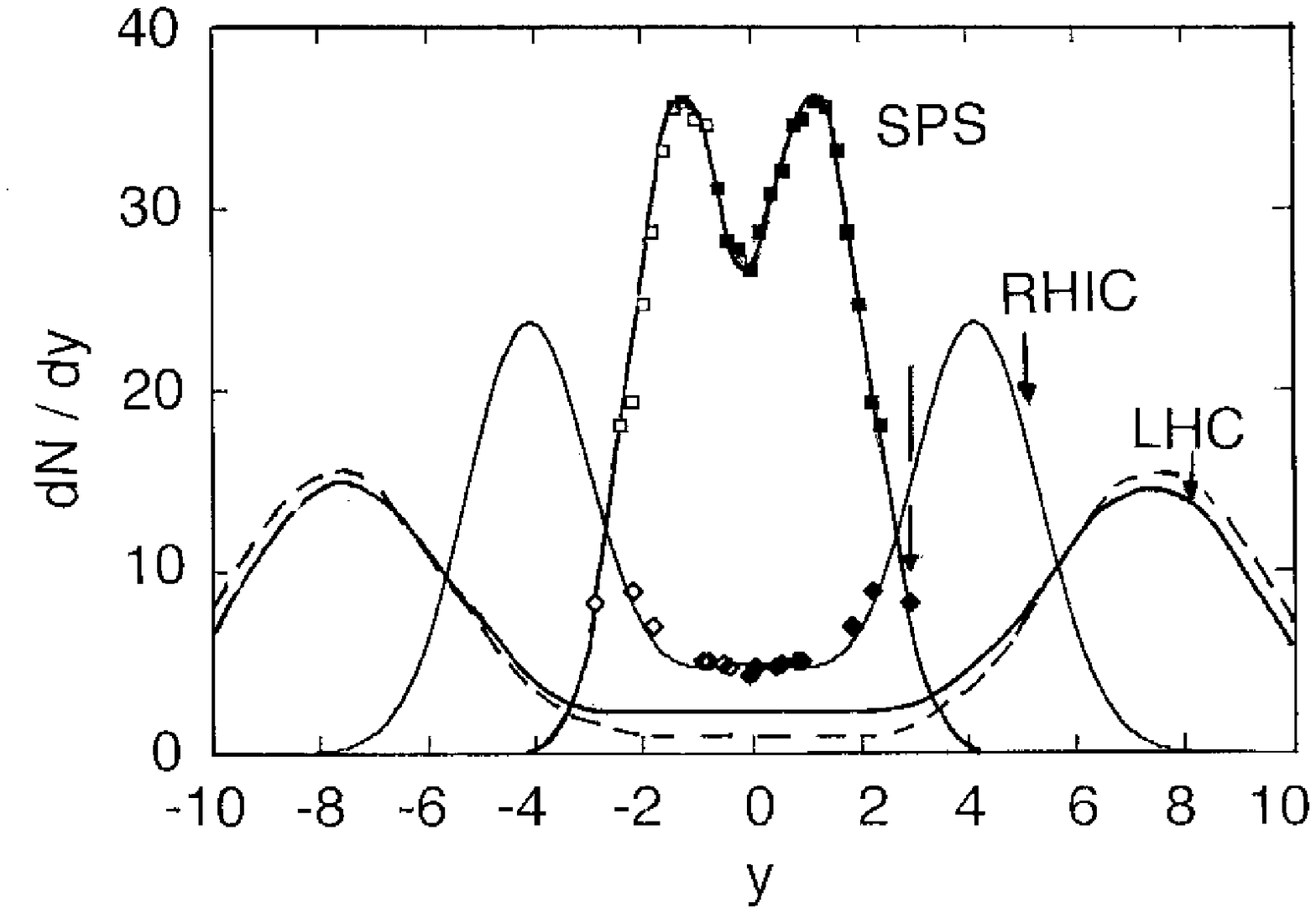}
\newpage
\vspace{1cm}
\includegraphics[width=16cm]{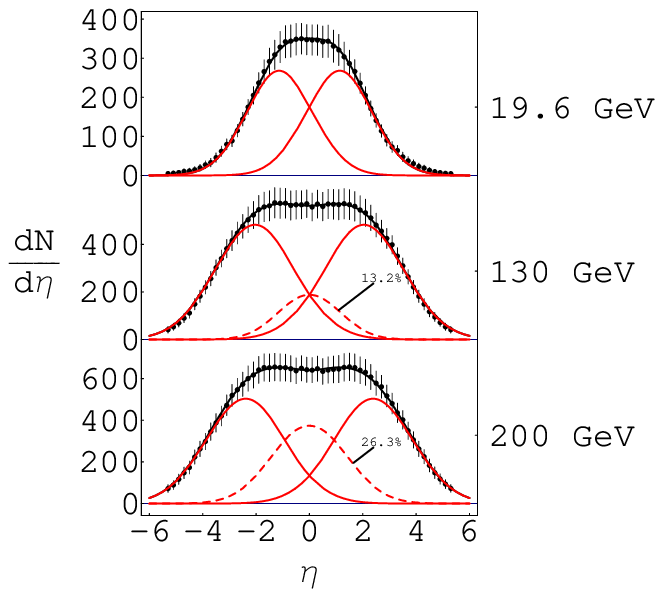}
\includegraphics[width=12cm]{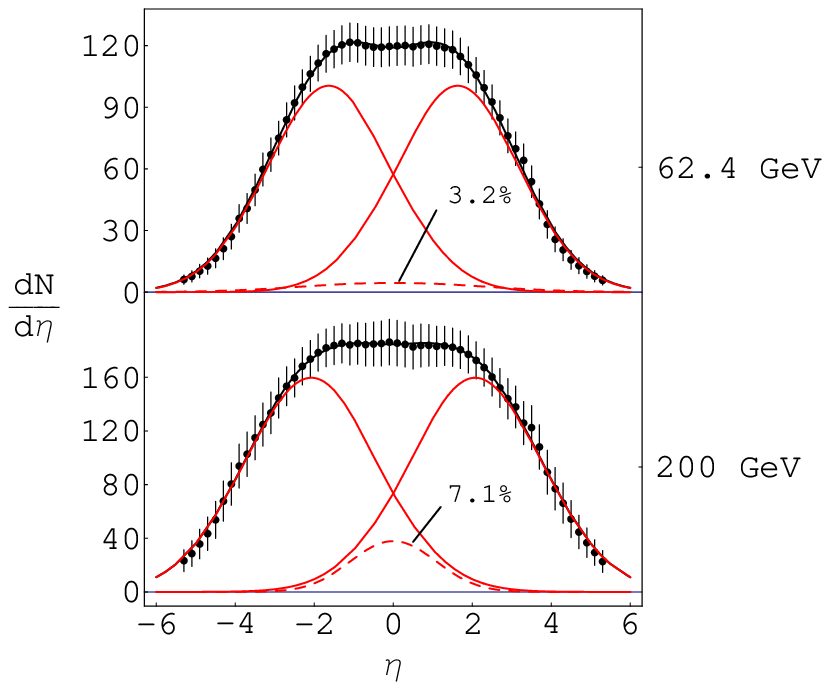}
\includegraphics[width=12cm]{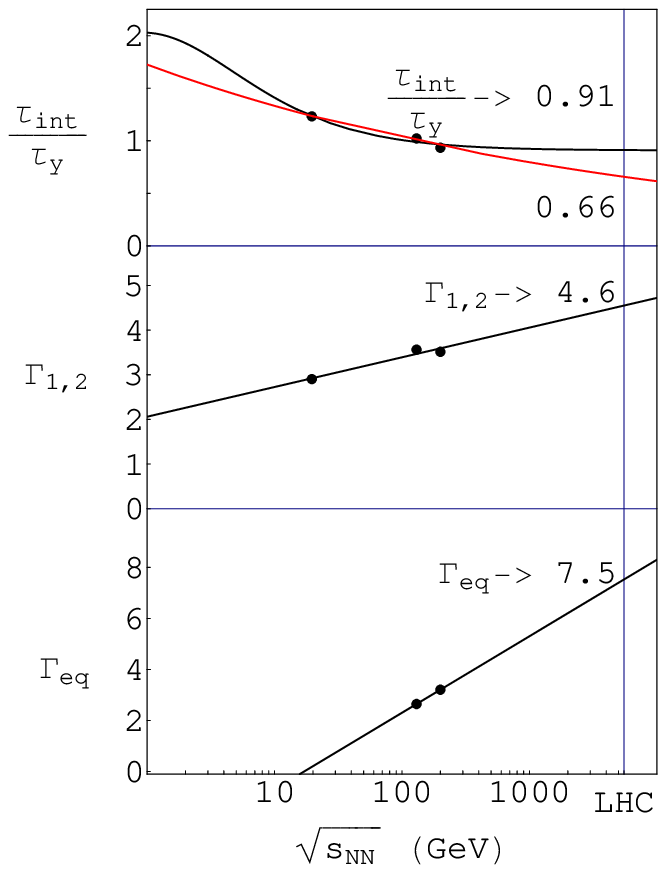}
\includegraphics[width=16cm]{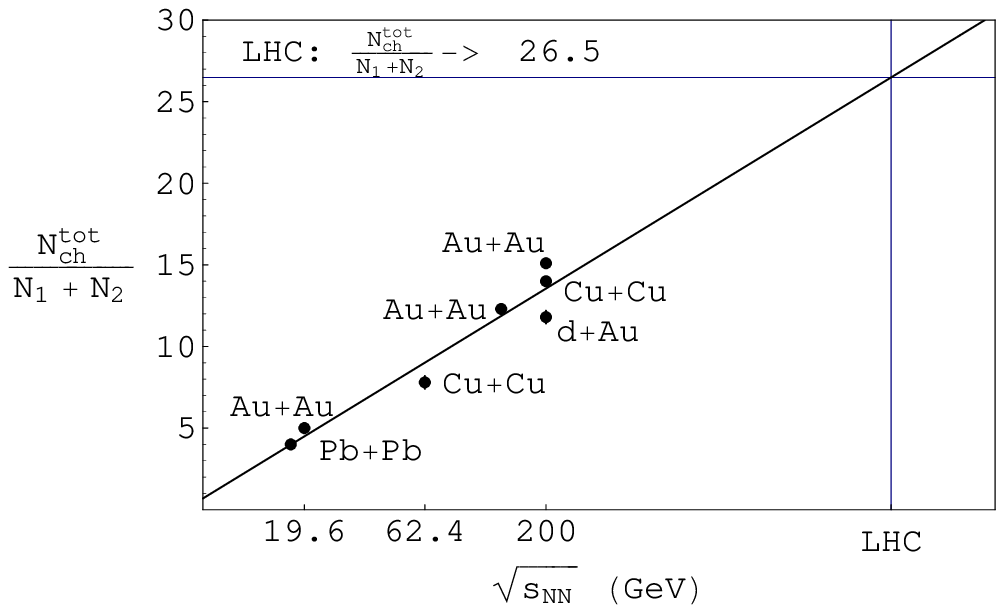}
\includegraphics[width=12cm]{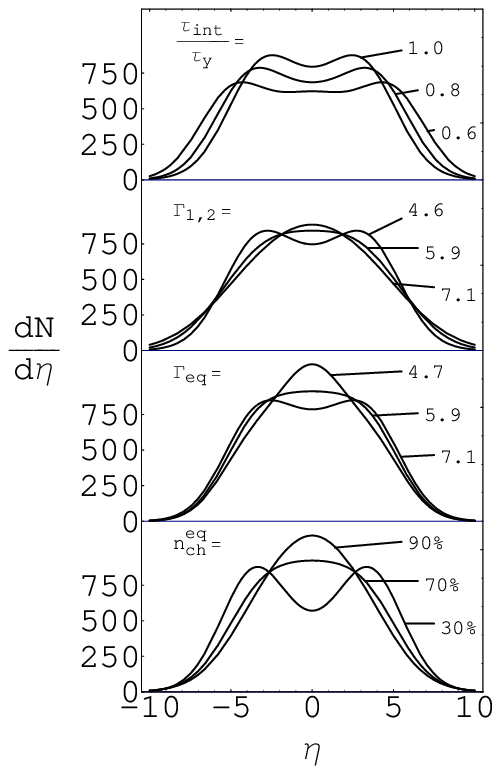}
\includegraphics[width=15.2cm]{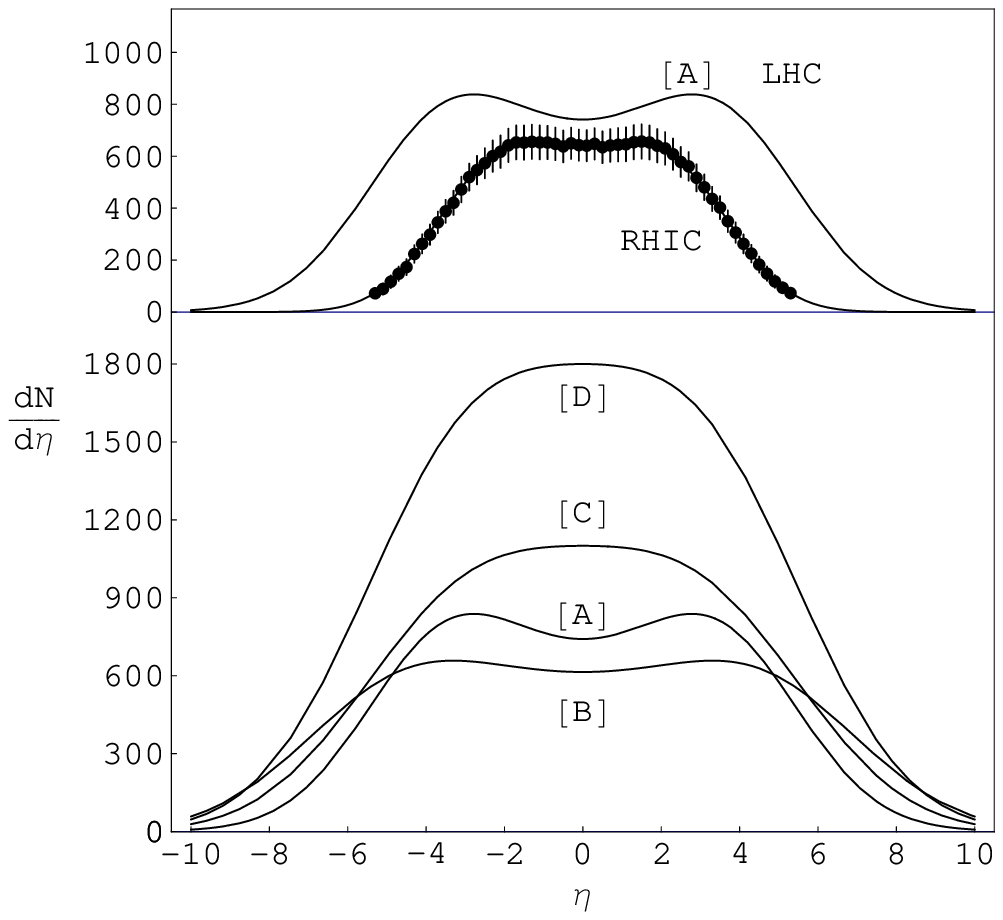}
\end{document}